\documentclass[%
reprint,
superscriptaddress,
 amsmath,amssymb,
 aps,
 prl,
 showkeys,
floatfix
]{revtex4-2}

\usepackage{graphicx}
\usepackage{subfigure}
\usepackage{dcolumn}
\usepackage{bm}
\usepackage{amsmath}
\usepackage{comment}
\usepackage{multirow}
\usepackage[colorlinks=true,      
            linkcolor=blue,       
            citecolor=blue,      
            urlcolor=blue      
           ]{hyperref}
\usepackage{cleveref}
\crefname{equation}{Equation}{Eqs.}
\crefname{figure}{Figure}{Figures}
\usepackage{url}
\usepackage{float}

\def\Qt2{{Q}^2}
\def\MM2{M_X^2}
\def\Jp{J$/\psi$~}

\begin{document}

\preprint{JLAB-XX-XXX-XXXX}
\title{First Measurement of Near-Threshold J$/\psi$ Photoproduction on the Neutron}
\newcommand*{\GLASGOW}{University of Glasgow, Glasgow G12 8QQ, United Kingdom}
\newcommand*{\GLASGOWindex}{40}
\affiliation{\GLASGOW}

\newcommand*{\SACLAY}{IRFU, CEA, Universit'{e} Paris-Saclay, F-91191 Gif-sur-Yvette, France}
\newcommand*{\SACLAYindex}{2}
\affiliation{\SACLAY}

\newcommand*{\JLAB}{Thomas Jefferson National Accelerator Facility, Newport News, Virginia 23606}
\newcommand*{\JLABindex}{34}
\affiliation{\JLAB}


\newcommand*{\ANL}{Argonne National Laboratory, Argonne, Illinois 60439}
\newcommand*{\ANLindex}{1}
\affiliation{\ANL}
\newcommand*{\CNU}{Christopher Newport University, Newport News, Virginia 23606}
\newcommand*{\CNUindex}{3}
\affiliation{\CNU}
\newcommand*{\UCONN}{University of Connecticut, Storrs, Connecticut 06269}
\newcommand*{\UCONNindex}{4}
\affiliation{\UCONN}
\newcommand*{\DUKE}{Duke University, Durham, North Carolina 27708-0305}
\newcommand*{\DUKEindex}{5}
\affiliation{\DUKE}
\newcommand*{\DUQUESNE}{Duquesne University, 600 Forbes Avenue, Pittsburgh, PA 15282 }
\newcommand*{\DUQUESNEindex}{6}
\affiliation{\DUQUESNE}
\newcommand*{\FU}{Fairfield University, Fairfield CT 06824}
\newcommand*{\FUindex}{7}
\affiliation{\FU}
\newcommand*{\FERRARAU}{Universita' di Ferrara , 44121 Ferrara, Italy}
\newcommand*{\FERRARAUindex}{8}
\affiliation{\FERRARAU}
\newcommand*{\FIU}{Florida International University, Miami, Florida 33199}
\newcommand*{\FIUindex}{9}
\affiliation{\FIU}
\newcommand*{\Genova}{Universit$\grave{a}$ di Genova, 16146 Genova, Italy}
\newcommand*{\Genovaindex}{10}
\affiliation{\Genova}
\newcommand*{\GWUI}{The George Washington University, Washington, DC 20052}
\newcommand*{\GWUIindex}{11}
\affiliation{\GWUI}
\newcommand*{\GSIFFN}{GSI Helmholtzzentrum fur Schwerionenforschung GmbH, D-64291 Darmstadt, Germany}
\newcommand*{\GSIFFNindex}{12}
\affiliation{\GSIFFN}
\newcommand*{\ORSAY}{Universit'{e} Paris-Saclay, CNRS/IN2P3, IJCLab, 91405 Orsay, France}
\newcommand*{\ORSAYindex}{13}
\affiliation{\ORSAY}
\newcommand*{\INFNCAT}{INFN, Sezione di Catania, 95123 Catania, Italy }
\newcommand*{\INFNCATindex}{14}
\affiliation{\INFNCAT}
\newcommand*{\INFNFE}{INFN, Sezione di Ferrara, 44100 Ferrara, Italy}
\newcommand*{\INFNFEindex}{15}
\affiliation{\INFNFE}
\newcommand*{\INFNFR}{INFN, Laboratori Nazionali di Frascati, 00044 Frascati, Italy}
\newcommand*{\INFNFRindex}{16}
\affiliation{\INFNFR}
\newcommand*{\INFNGE}{INFN, Sezione di Genova, 16146 Genova, Italy}
\newcommand*{\INFNGEindex}{17}
\affiliation{\INFNGE}
\newcommand*{\INFNRO}{INFN, Sezione di Roma Tor Vergata, 00133 Rome, Italy}
\newcommand*{\INFNROindex}{18}
\affiliation{\INFNRO}
\newcommand*{\INFNTUR}{INFN, Sezione di Torino, 10125 Torino, Italy}
\newcommand*{\INFNTURindex}{19}
\affiliation{\INFNTUR}
\newcommand*{\INFNPAV}{INFN, Sezione di Pavia, 27100 Pavia, Italy}
\newcommand*{\INFNPAVindex}{20}
\affiliation{\INFNPAV}
\newcommand*{\KNU}{Kyungpook National University, Daegu 41566, Republic of Korea}
\newcommand*{\KNUindex}{21}
\affiliation{\KNU}
\newcommand*{\LAMAR}{Lamar University, 4400 MLK Blvd, PO Box 10046, Beaumont, Texas 77710}
\newcommand*{\LAMARindex}{22}
\affiliation{\LAMAR}
\newcommand*{\MIT}{Massachusetts Institute of Technology, Cambridge, Massachusetts  02139-4307}
\newcommand*{\MITindex}{23}
\affiliation{\MIT}
\newcommand*{\MISS}{Mississippi State University, Mississippi State, MS 39762-5167}
\newcommand*{\MISSindex}{24}
\affiliation{\MISS}
\newcommand*{\UNH}{University of New Hampshire, Durham, New Hampshire 03824-3568}
\newcommand*{\UNHindex}{25}
\affiliation{\UNH}
\newcommand*{\NMSU}{New Mexico State University, PO Box 30001, Las Cruces, NM 88003, USA}
\newcommand*{\NMSUindex}{26}
\affiliation{\NMSU}
\newcommand*{\OHIOU}{Ohio University, Athens, Ohio  45701}
\newcommand*{\OHIOUindex}{27}
\affiliation{\OHIOU}
\newcommand*{\ODU}{Old Dominion University, Norfolk, Virginia 23529}
\newcommand*{\ODUindex}{28}
\affiliation{\ODU}
\newcommand*{\JLUGiessen}{II Physikalisches Institut der Universitaet Giessen, 35392 Giessen, Germany}
\newcommand*{\JLUGiessenindex}{29}
\affiliation{\JLUGiessen}
\newcommand*{\URICH}{University of Richmond, Richmond, Virginia 23173}
\newcommand*{\URICHindex}{30}
\affiliation{\URICH}
\newcommand*{\ROMAII}{Universita' di Roma Tor Vergata, 00133 Rome Italy}
\newcommand*{\ROMAIIindex}{31}
\affiliation{\ROMAII}
\newcommand*{\SCAROLINA}{University of South Carolina, Columbia, South Carolina 29208}
\newcommand*{\SCAROLINAindex}{32}
\affiliation{\SCAROLINA}
\newcommand*{\TEMPLE}{Temple University,  Philadelphia, PA 19122 }
\newcommand*{\TEMPLEindex}{33}
\affiliation{\TEMPLE}
\newcommand*{\ULS}{Universidad de La Serena}
\newcommand*{\ULSindex}{35}
\affiliation{\ULS}
\newcommand*{\UTFSM}{Universidad T\'{e}cnica Federico Santa Mar\'{i}a, Casilla 110-V Valpara\'{i}so, Chile}
\newcommand*{\UTFSMindex}{36}
\affiliation{\UTFSM}
\newcommand*{\INSUBRIA}{Universit\`{a} degli Studi dell'Insubria, 22100 Como, Italy}
\newcommand*{\INSUBRIAindex}{37}
\affiliation{\INSUBRIA}
\newcommand*{\BRESCIA}{Universit`{a} degli Studi di Brescia, 25123 Brescia, Italy}
\newcommand*{\BRESCIAindex}{38}
\affiliation{\BRESCIA}
\newcommand*{\UCR}{University of California Riverside, 900 University Avenue, Riverside, CA 92521, USA}
\newcommand*{\UCRindex}{39}
\affiliation{\UCR}
\newcommand*{\YORK}{University of York, York YO10 5DD, United Kingdom}
\newcommand*{\YORKindex}{41}
\affiliation{\YORK}
\newcommand*{\VIRGINIA}{University of Virginia, Charlottesville, Virginia 22901}
\newcommand*{\VIRGINIAindex}{42}
\affiliation{\VIRGINIA}
\newcommand*{\WM}{College of William and Mary, Williamsburg, Virginia 23187-8795}
\newcommand*{\WMindex}{43}
\affiliation{\WM}
\newcommand*{\YEREVAN}{Yerevan Physics Institute, 375036 Yerevan, Armenia}
\newcommand*{\YEREVANindex}{44}
\affiliation{\YEREVAN}
\newcommand*{\MSU}{Skobeltsyn Nuclear Physics Institute, Lomonosov Moscow State University, 119234 Moscow, Russia}
\newcommand*{\MSUindex}{45}
\affiliation{\MSU}

\newcommand*{\NOWUCONN}{University of Connecticut, Storrs, Connecticut 06269}
\newcommand*{\NOWDOTA}{DOTA, ONERA, Universit\'{e} Paris-Saclay, 91120, Palaiseau, France}
\newcommand*{\NOWODU}{Old Dominion University, Norfolk, Virginia 23529}

\author{R.~Tyson}
\email{Corresponding author: {tyson@jlab.org}}
\affiliation{\GLASGOW}

\author{D.G.~Ireland}
\affiliation{\GLASGOW}

\author{B.~McKinnon}
\affiliation{\GLASGOW}

\author{P.~Chatagnon}
\affiliation{\SACLAY}

\author{V.~Kubarovsky}
\affiliation{\JLAB}

\author{S.~Stepanyan}
\affiliation{\JLAB}

\author {A.G.~Acar} 
\affiliation{\YORK}
\author {P.~Achenbach} 
\affiliation{\CNU}
\author {J.S.~Alvarado} 
\affiliation{\ORSAY}
\author {M.J.~Amaryan} 
\affiliation{\ODU}
% ...

\collaboration{The CLAS Collaboration}

\date{\today}
\begin{abstract}
We report the first measurement of the total and $t$-differential cross sections for near-threshold \Jp photoproduction on the neutron, obtained with the CLAS12 detector at the Thomas Jefferson National Accelerator Facility. The measurements, performed using a liquid-deuterium target, also provide the incoherent total and $t$-differential cross sections on the bound proton, enabling the first direct comparison of near-threshold \Jp photoproduction on bound protons and neutrons. Interpreted within Vector Meson Dominance, holographic QCD, and GPD-based frameworks, the data allow for the investigation of the gluonic structure of bound nucleons. A comparison with free proton results reveals hints of modifications to the gluon structure of nucleons in the nuclear medium, within the assumptions of the extraction and on the underlying production mechanism. The comparison of bound neutron and bound proton cross sections provides new constraints on the near-threshold \Jp production mechanism, which is crucial in order to establish \Jp photoproduction as a probe of the nucleon's gluonic structure.
\end{abstract}
\keywords{CLAS12, J$/\psi$, photoproduction, neutron, gluons}
\maketitle

\section{Introduction}
Heavy vector meson photoproduction is regarded as a key tool for studying the gluonic degrees of freedom within nucleons. In recent years, the Thomas Jefferson National Accelerator Facility (JLab) GlueX~\cite{Gluex1,GlueX2}, J$/\psi$-007~\cite{HallC1,HallC2}, and CLAS12~\cite{CLAS} Collaborations reported measurements of \Jp photoproduction on the free proton. These measurements have been related to the proton's gluon content, with Vector Meson Dominance (VMD)-based models providing a new avenue to probe nucleon structure~\cite{VMD1,VMD2,VMD3}. In these models, the predicted dominance of two-gluon exchange near the photoproduction threshold relates the \Jp photoproduction cross section to the proton Energy Momentum Tensor (EMT)~\cite{EMT1,EMT2}. This relationship enables the extraction of nucleon properties encoded in the gluon mechanical form factors~\cite{GFFs1,GFFs3}, including the trace anomaly contribution to the proton mass~\cite{Trace0,Trace1,Trace2} and the proton mass radius~\cite{MassRad,MassRad2}. Extractions of the \Jp-proton scattering length provide constraints on possible \Jp-proton bound states and the proton’s gluonic structure~\cite{ScatLength,ScatLength2}.

A range of complementary theoretical frameworks have also been proposed, notably holographic QCD~\cite{Holo1,Holo2,Holo3,Holo4} and generalized parton distribution (GPD) models~\cite{GPD1,GPD2,GPD3,GPD4}. In holographic QCD, based on gauge/gravity duality constructions such as the AdS/CFT correspondence, the nucleon's non-perturbative structure is encoded in higher-dimensional gravitational dynamics. GPD-based approaches parameterize correlated parton momentum and spatial distributions. Both frameworks provide access to gluon distributions through exclusive processes such as near-threshold \Jp photoproduction.

The GlueX, J$/\psi$-007, and CLAS12 data have subsequently been used to extract the free-proton gluon mechanical form factors $A_g(t)$ and $C_g(t)$, and their momentum-transfer dependence~\cite{HallC1,HallCGFF,HallC2,CLAS}. The $A_g(t)$ form factor characterizes how gluons carry and spatially distribute momentum and energy within the nucleon, while $C_g(t)$ describes their contribution to internal pressure and shear forces. These form factors also determine effective gluonic radii, including a mass radius associated with the spatial distribution of gluonic energy and a scalar radius related to the trace of the energy--momentum tensor. Overall, these extractions~\cite{HallC1,HallC2} agree within uncertainties with lattice QCD predictions~\cite{LQCD1,LQCD2,LQCD3}.

While the gluonic structure of the free proton has been extensively studied, much less is known about bound nucleons in nuclei. The Short-Range Correlation-Color Transparency (SRC-CT) experiment measured near- and sub-threshold \Jp production on protons in deuterium, helium, and carbon~\cite{SRCCT}. The observed enhancement of the sub-threshold cross section could indicate modified gluon structure in bound protons~\cite{SRCCT,SubTh,ScatLengthNuclei}. However, the statistical precision of these data limits the strength of this conclusion, motivating a follow-up experiment approved in 2024~\cite{SRCCTProp}.

Near-threshold \Jp production is generally assumed at leading order to proceed via two-gluon exchange. However, alternative models include non-negligible contributions from three-gluon exchange~\cite{Gluex1,VMD2}, open-charm intermediate states~\cite{OpenCharm1,OpenCharm2}, Pomeron~\cite{POM1,POM2} or meson~\cite{MesonExchange} exchange, and hidden-charm pentaquarks observed by LHCb in 2015 and 2019~\cite{Penta1,Penta2,Penta3,Penta4,IgorPenta}. Understanding the production mechanism is crucial, as only gluonic exchange enables \Jp photoproduction to probe the gluon structure of the target nucleon. A detailed analysis by the JPAC Collaboration examined the contributions of these mechanisms to the existing free-proton cross section data~\cite{JPAC}. However, no mechanism was conclusively excluded, highlighting the need for additional data.

This Letter presents measurements of the total and $t$-differential cross sections for near-threshold \Jp quasi-real photoproduction on the bound neutron and bound proton with the CLAS12 experiment using a liquid-deuterium target. This constitutes the first measurement on the bound neutron. Two-gluon exchange is isospin invariant, whereas open-charm contributions to the neutron and proton channels proceed through different intermediate states. Comparing the bound-neutron and bound-proton cross sections therefore constrains the near-threshold \Jp production mechanism. Holographic QCD and GPD-based models have been used to describe the neutron gluon structure. By combining the new neutron and proton CLAS12 cross sections, the gluonic structure of bound nucleons has been investigated, suggesting possible modifications to the gluon structure in the nuclear medium.

\section{The experiment}
\label{sec:CLAS12}

The Continuous Electron Beam Accelerator Facility~(CEBAF)~\cite{CEBAF} at JLab delivers a polarized electron beam to the four experimental halls. The data analyzed in this work were collected with the CEBAF Large Acceptance Spectrometer at 12~GeV (CLAS12) in Hall~B. CLAS12 comprises a forward detector covering polar angles from 5$^{\circ}$ to 35$^{\circ}$ and a central detector covering 35$^{\circ}$ to 125$^{\circ}$, both with nearly full azimuthal acceptance~\cite{CLAS12}. In 2019 and 2020, CLAS12 collected data using a 5-cm-long liquid-deuterium (LD$_2$) target, enabling the bound neutron and bound proton measurements presented here. Three datasets were recorded with beam energies of 10.2~GeV, 10.4~GeV, and 10.6~GeV and integrated luminosities of 60~fb$^{-1}$, 41~fb$^{-1}$, and 62~fb$^{-1}$, respectively. Apart from the beam energy, all datasets were taken under identical conditions with beam currents of 40 to 55~nA. The datasets were combined to maximize the \Jp yield, as the resolution of the key reconstructed quantities was consistent across all datasets.

This analysis followed a strategy similar to that of Ref.~\cite{CLAS}, which measured the \Jp production on the free proton. The scattered electron was not required for the analysis, enabling access very low quasi-real photon virtuality. The \Jp mesons were identified from the invariant mass of their decay electron and positron detected alongside the recoil nucleon (either neutron or proton) in the CLAS12 forward detector.  The CLAS12 trigger recorded events containing a forward electron with an efficiency exceeding 99.5\%~\cite{Trigger}. The CLARA and COATJAVA frameworks~\cite{COATJAVA} reconstructed and identified particles using signals from multiple detector subsystems, while CLAS12ROOT and CHANSER were used for further analysis~\cite{CLAS12ROOT,CHANSER}. The drift chambers~\cite{DC} measured the momenta and angles of charged particles bent by the CLAS12 torus magnet~\cite{Magnets}. The forward time-of-flight system identified charged hadrons using the mass hypothesis best matching the measured flight time~\cite{FTOF}. Below 4.5~GeV, the high-threshold Cherenkov counters vetoed charged pions to identify electrons and positrons~\cite{HTCC}. Over the full momentum range, particle energy deposition and shower profiles in the forward electromagnetic calorimeter separated electrons and positrons from charged pions~\cite{ECAL}. A boosted decision tree implemented with the ROOT TMVA framework~\cite{TMVA,Voss:2007jxm} further improved pion rejection~\cite{CLAS}. The calorimeter also detected neutral particles. The neutron detection efficiency was measured from the ratio of $ep \to e'\pi^+n$ to $ep \to e'\pi^+(n)$ events, where the undetected neutron $(n)$ was expected to intersect the calorimeter fiducial volume. Corrections were applied to the final-state particle momenta, and electrons and positrons detected near calorimeter edges, where showers could leak outside the active volume, were rejected.

To select quasi-real photoproduction of electron-positron pairs, the missing mass in the reaction $eN \to e^+ e^-N'X$ was required to be consistent with the undetected scattered electron, where $N$ ($N'$) denotes the interacting bound (recoil) nucleon and $X$ the missing particles. The quasi-real photon virtuality, \mbox{$Q^{2}=2E_{\text{beam}}E_{\text{X}}(1-\cos\theta_{\text{X}})$}, and energy, $E_{\gamma}=E_{\text{beam}}-E_{\text{X}}$, were reconstructed from the missing-particle kinematics, where $E_{\rm beam}$ is the beam energy and $E_{\text{X}}$ and $\theta_{\text{X}}$ are the missing-particle energy and polar angle. The momentum transfer was calculated as $t=(\mathbf{N'}-\mathbf{N})^2$ using the recoil and target nucleon four-momenta. The bound nucleon was assumed to be at rest, neglecting Fermi motion, which smears the resolution on quantities such as $Q^2$ and $t$. Its effect was minimized in the photon-energy reconstruction following the prescription of Ref.~\cite{JpsiD_SOLID}.

\begin{figure}
\centering
\includegraphics[width=\linewidth]{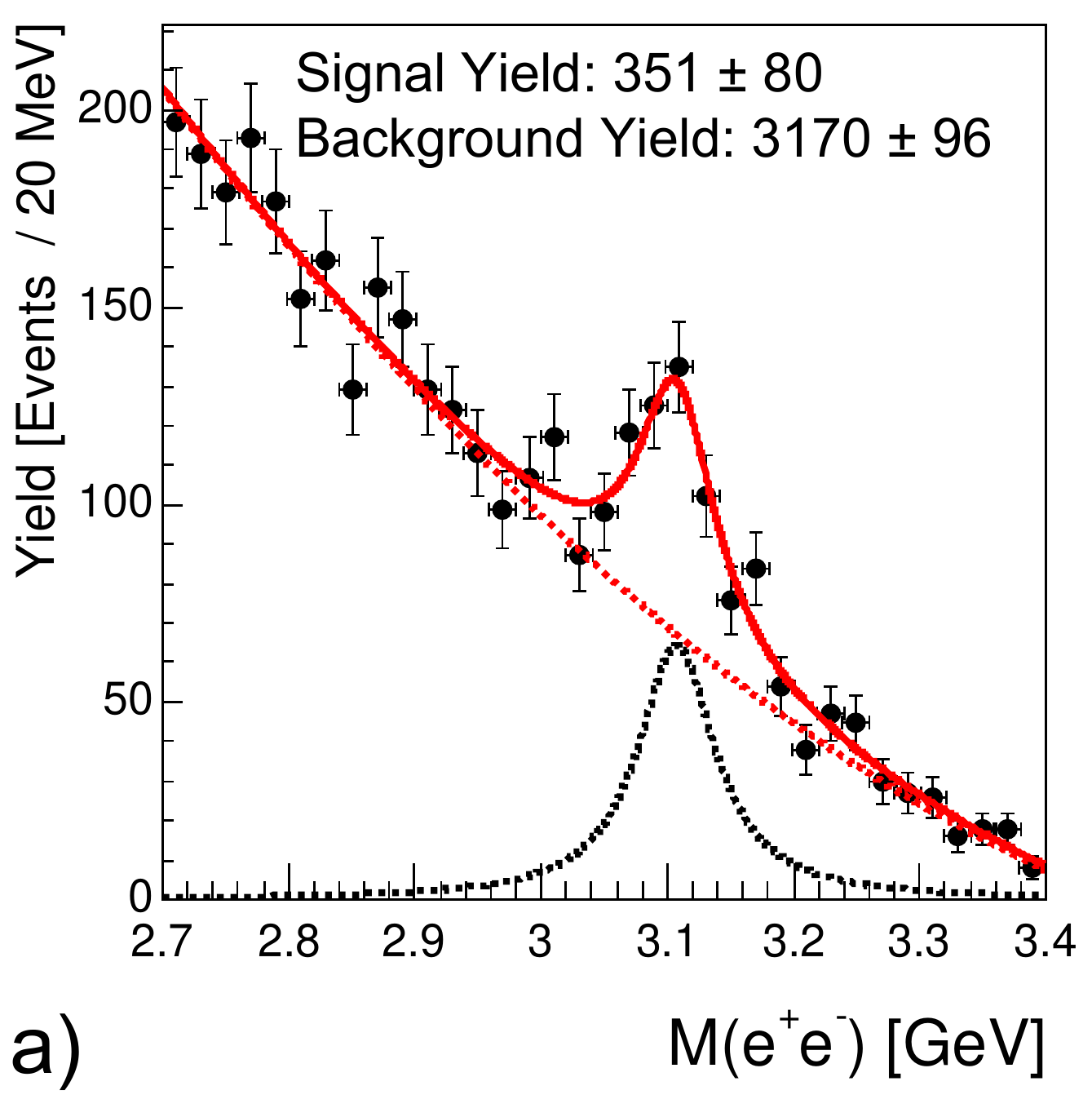}\\
\includegraphics[width=\linewidth]{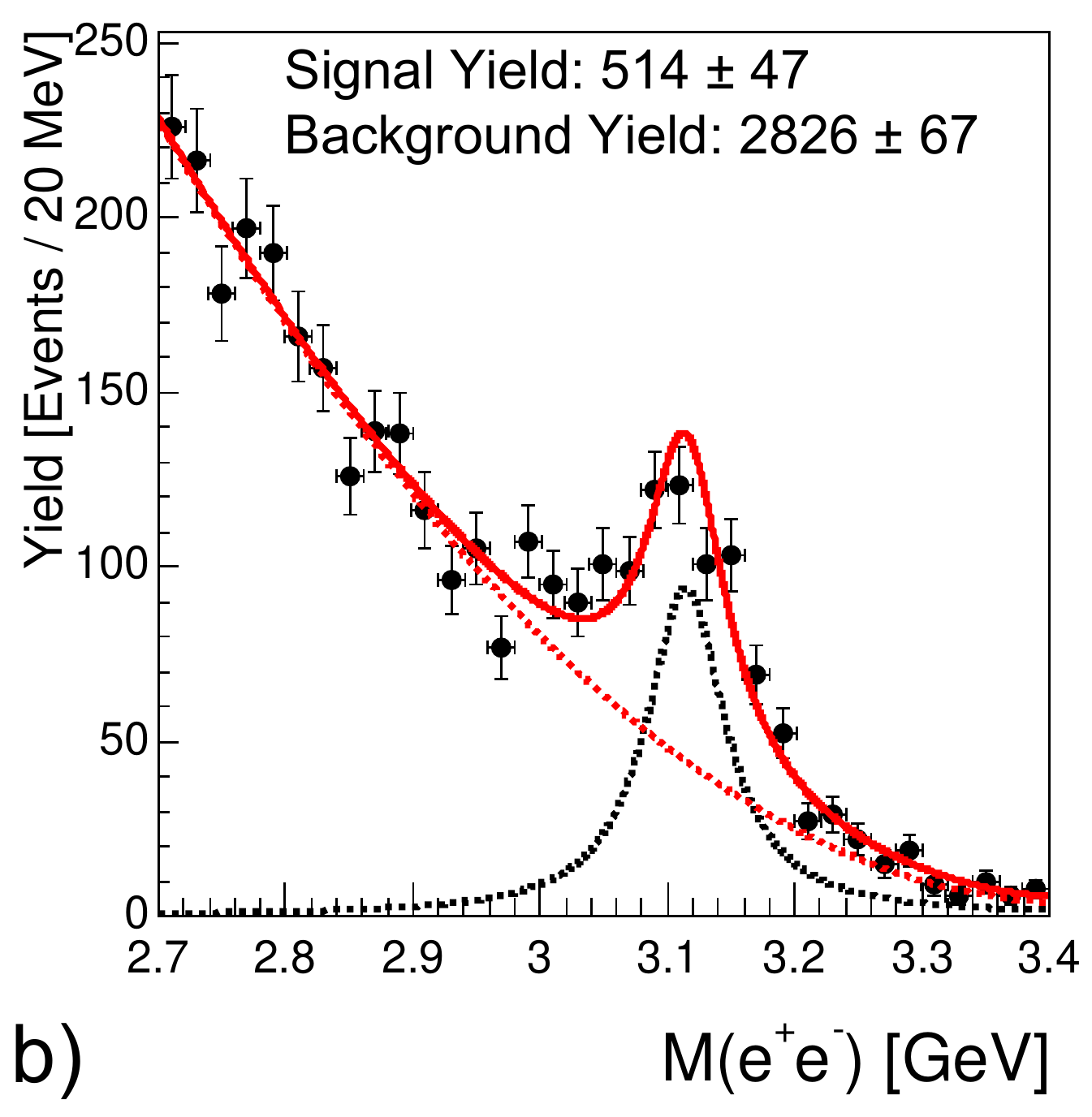}
\caption{The $e^{+}e^{-}$ invariant mass distribution $M(e^{+}e^{-})$ produced over the full photon energy and $t$ range on: a) bound neutron, b) bound proton. The distributions are fitted with a second-order polynomial for the background and a non-relativistic Breit-Wigner for the \Jp signal.}
\label{art:fig_fit}
\end{figure}

Figure~\ref{art:fig_fit} shows \Jp production over the full photon-energy and $t$ ranges for the neutron and proton, tagged by detecting a proton or neutron in the forward detector and satisfying the missing-mass requirement for $eN_{\text{bound}} \to e^+ e^-N'X$. The electron-positron invariant mass was fitted to extract the \Jp yield using a Breit-Wigner function for the signal and a second-order polynomial for the background. The choice of fit functions was validated with thousands of pseudo-experiments, in which artificial invariant-mass spectra with known \Jp and background yields were generated and fitted. The chosen functions minimized the average difference between the injected and extracted \Jp yields.

The total cross section as a function of $E_{\gamma}$ and the $t$-differential cross section were calculated using Eqs.~(\ref{CalcXS}) and~(\ref{CalcDXS}), respectively.

\begin{equation}\label{CalcXS}
    \sigma_{0}(E_{\gamma})=\frac{N_{{\rm J}/\psi}(E_{\gamma})}{N_{\gamma}\,\rho_{T}\,l_{T}\,\omega_{c}R_{c}(E_{\gamma})\, B_r\,\epsilon(E_{\gamma})},
\end{equation}

\begin{equation}\label{CalcDXS}
    \frac{d\sigma(E_{\gamma})}{dt}=\frac{N_{{\rm J}/\psi}(t,E_{\gamma})}{N_{\gamma}\,\rho_{T}\, l_{T}\,\omega_{c}\, R_{c}(E_{\gamma})\, B_r\,\epsilon(t,E_{\gamma})\,\Delta t}.
\end{equation}

Here, $N_{{\rm J}/\psi}$ is the fitted \Jp yield in bins of $E_{\gamma}$ or $E_{\gamma}$ and $t$. The photon flux $N_{\gamma}$ was determined from the accumulated beam charge on the $\text{LD}2$ target, the Equivalent Photon Approximation virtual photon flux, and the real photon flux from bremsstrahlung photons produced in the target and target cell~\cite{EPA,EPAFlux,PDG}. The target density and length are denoted by $\rho_{T}$ and $l_{T}$, respectively. $R_{c}$ accounts for radiative effects~\cite{CLAS,RadCor6}, $B_r=0.0597$ is the \Jp branching ratio to electron-positron pairs~\cite{PDG}, and $\Delta t$ accounts for the $t$-bin width. The factor $\epsilon$ represents the detector acceptance and efficiency, while $\omega_c$ corrects $\epsilon$ and the photon flux following the prescription of Refs.~\cite{Gluex1,GlueX2,CLAS}.

The efficiency $\epsilon$ was determined using simulated \Jp events generated with JPsiGen~\cite{JPsiGen} and propagated through the Geant4-based CLAS12 simulation GEMC~\cite{GEMC}. Random-trigger background events were added to emulate physics and electronic backgrounds, allowing their impact on reconstruction efficiency and resolution to be reproduced. Simulated momenta and angles were smeared to match the detector resolution observed in CLAS12 data. A combinatorial background was constructed by combining final-state particles from different events and added to the simulated electron-positron invariant mass spectrum. The resulting distributions were fitted using the same signal and background parameterization as the data, incorporating any fitting bias into the efficiency determination.

The dominant non-resonant electron-positron production mechanism is the Bethe-Heitler~(BH) process, whose cross section is precisely calculable in quantum electrodynamics. This cross section was used to simulate BH electron-positron production in CLAS12~\cite{TCSGen,BHDeut}. Reference~\cite{CLAS} demonstrated that BH events can be identified in CLAS12 data through a peak at $Q^2=0$~GeV$^2$ for events with invariant masses near the \Jp mass.

The correction factor $\omega_c$ was calculated as the ratio of simulated BH events to those observed in CLAS12 data, with deviations from unity reflecting residual differences between simulated and true detector efficiencies. Corrections were applied to the simulated rate of change of the efficiency as a function of beam current, electron and positron identification efficiencies, neutron detection efficiency, and $Q^{2}$ and missing-mass cut efficiencies. After applying these corrections, $\omega_c=0.954\pm0.193$. Final-state interaction contributions to the BH process were estimated to be negligible compared with the statistical uncertainty on $\omega_c$~\cite{BHDeut}.

The systematic uncertainty was evaluated by varying analysis procedures for which the resulting changes exceeded statistical fluctuations at the 95\% confidence level, following the Barlow significance criterion~\cite{Barlow}. The uncertainty from the fitting procedure was evaluated using alternative signal and background parameterizations and was small because fit-induced biases were already included through $\epsilon$. Additional contributions arose from uncertainties in radiative corrections, accumulated charge, and the statistical uncertainty on $\omega_c$, which dominates the total uncertainty. The total systematic uncertainty was approximately 25 to 30\% of the cross section, with $\omega_c$ contributing 20\%. For comparison, the statistical uncertainty ranged from 20 to 30\%.

\section{Results and interpretation}
\label{sec:Results}

\begin{figure}[hbtp]
    \centering   
    \includegraphics[width=\linewidth]{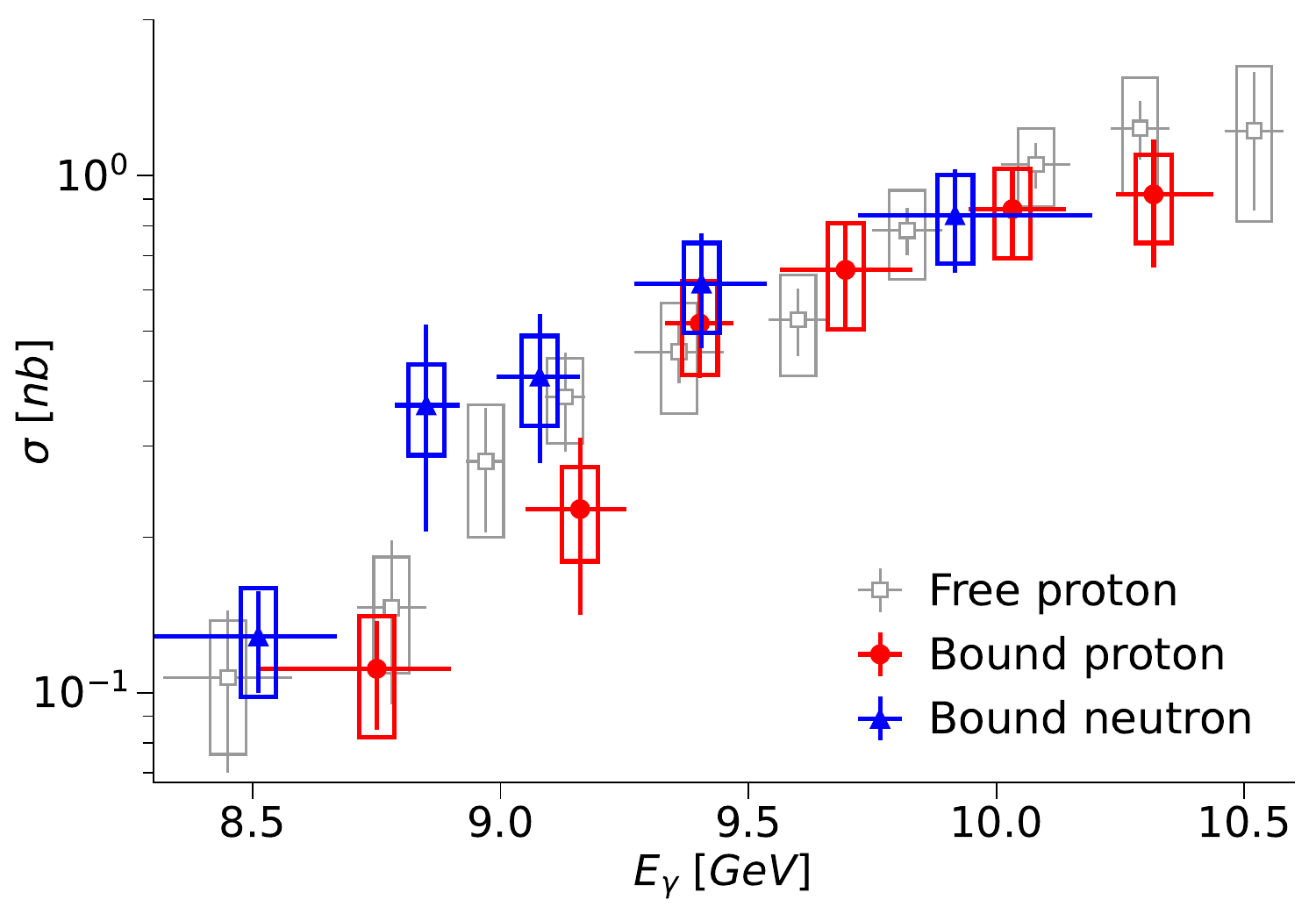}
    \caption[The \Jp photoproduction cross section.]{
    Total cross section for \Jp photoproduction as a function of $E_{\gamma}$ on the bound proton~(red circles) and neutron~(blue triangles), compared with the free-proton CLAS12 measurement~\cite{CLAS}~(gray). Vertical error bars represent statistical uncertainties and boxes represent systematic uncertainties. Data points are plotted at the mean $E_{\gamma}$ value in each bin, with horizontal error bars showing the central 68\% interval of the sampled $E_{\gamma}$ distribution within each bin.}
    \label{art:xsect_tot}
\end{figure}

\begin{figure}[hbtp]
    \centering   
    \includegraphics[width=\linewidth]{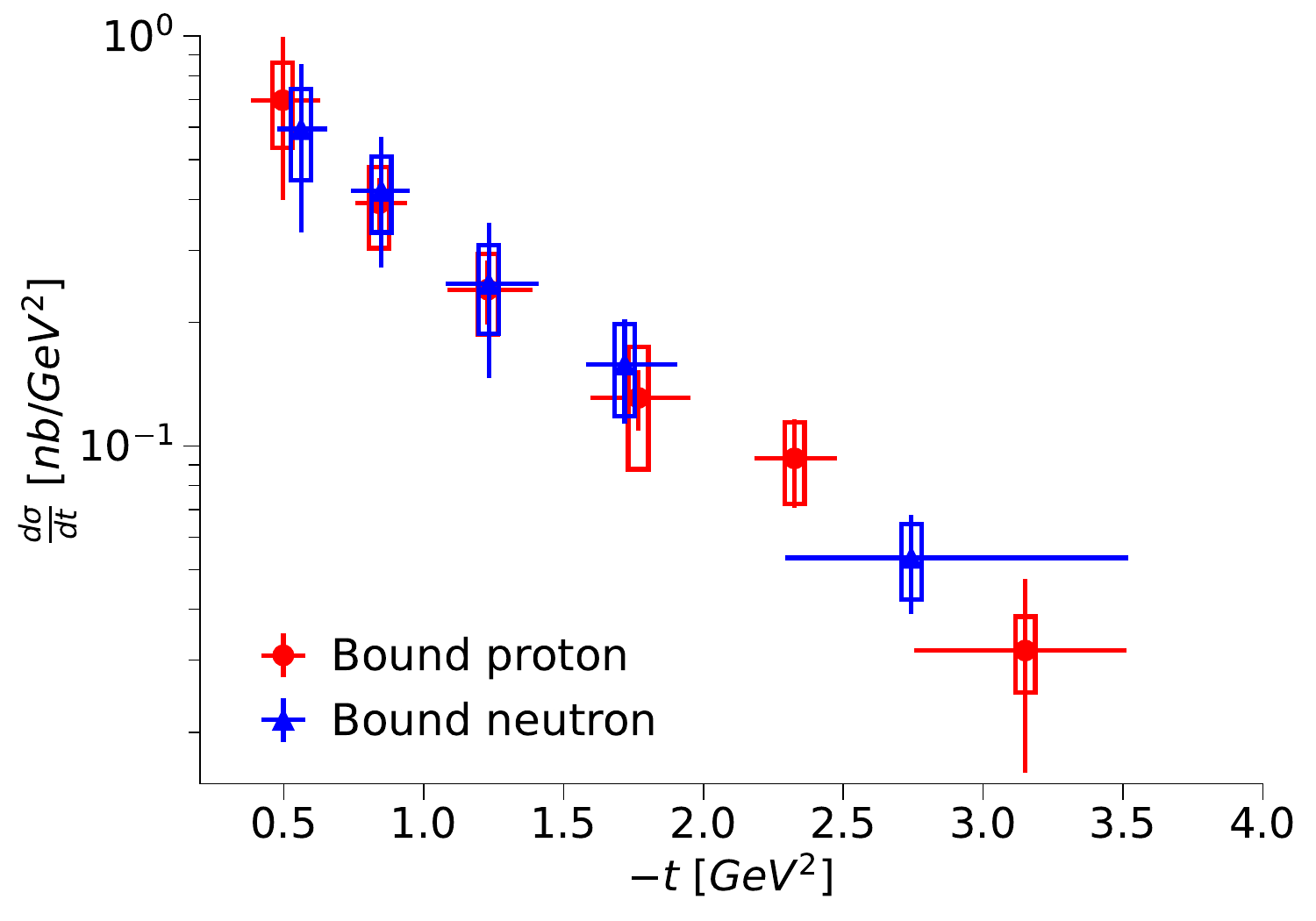}
    \caption[The \Jp photoproduction differential cross section.]{Differential cross section for \Jp photoproduction as a function of $-t$ for the bound proton (red circles) and neutron (blue triangles) in the range 8.2$<$$E_{\gamma}$$<$10.6~GeV, using the same conventions as Fig.~\ref{art:xsect_tot}.}
    \label{art:xsect_dif}
\end{figure}

The total and differential cross sections are shown in Figs.~\ref{art:xsect_tot} and~\ref{art:xsect_dif}, respectively, with numerical results reported in the End Matter. The cross sections were evaluated at the mean $E_{\gamma}$ or $t$ value in each bin. The horizontal bars indicate the central 68\% interval of the sampled $E_{\gamma}$ or $t$ distribution within each bin, defined by the 16th and 84th percentiles. The asymmetric intervals account for the non-uniform event weighting arising from the energy dependence of the photon flux and cross section. Measurements from GlueX~\cite{Gluex1,GlueX2}, J$/\psi$-007~\cite{HallC1,HallC2}, and CLAS12~\cite{CLAS} are compatible within uncertainties with the bound neutron and proton results presented here. The agreement between free and bound proton cross sections indicates that any final-state interaction contribution is smaller than the statistical and systematic uncertainties. Similarly, the agreement between bound neutron and proton cross sections suggests that any violation of isospin invariance in the \Jp production mechanism is below the sensitivity of this measurement.

Due to limited statistics, the differential cross section was extracted in a single $E_{\gamma}$ bin from 8.2 to 10.6~GeV. As shown in Fig.~\ref{art:xsect_dif}, the neutron and proton cross sections are compatible within uncertainties. The differential cross sections include contributions from quasi-free scattering on the neutron or proton of the deuteron target, as well as final-state interactions involving the scattered nucleons and the produced \Jp with spectator nucleons. These contributions were predicted to be several orders of magnitude smaller than the quasi-free contribution~\cite{FSI,FSI2} and are therefore neglected. This assumption corroborates the agreement within uncertainties of the total cross sections on bound and free protons.

The nucleon mass radius was extracted using a dipole fit following the VMD-based prescription of Ref.~\cite{MassRad}, assuming negligible final-state interaction contributions. The bound neutron and proton data were combined to extract the nucleon mass radius. The extracted values are given in Table~\ref{art:tab_mrGFF}. The bound neutron and proton mass radii are compatible within uncertainties, indicating similar gluonic structures.

The nucleon mechanical form factors are defined through the EMT matrix elements~\cite{GFFs1} as:
\begin{align}
&\langle p_f, s_f | T_{q,g}^{\mu,\nu}(0) | p_i,s_i \rangle = \\  \nonumber
&\bar u(p_f,s_f) \displaystyle{\Bigl ( } A_{q,g}(t) \gamma^{ \{ \mu} P^{\nu \} }  + B_{q,g} \frac{iP^{ \{ \mu }\sigma^{\nu \} \rho}\Delta_{\rho}}{2M_N} \\  \nonumber
&
+ C_{g,q} \frac{\Delta^{\mu}\Delta^{\nu} -g^{\mu\nu}\Delta^2}{M_N} + \bar C_{q,g}(t) M_N g^{\mu,\nu} \displaystyle{ \Bigr )} u(p_i,s_i),
\end{align}
where $A_{q,g}(t)$, $B_{q,g}(t)$, $C_{q,g}(t)$, and $\bar C_{q,g}(t)$ are the quark and gluon form factors associated with the EMT components $T_{q,g}^{\mu,\nu}$. Here, $p_i$ ($p_f$) and $s_i$ ($s_f$) denote the initial (final) nucleon four-momenta and spin states, $u(p,s)$ is the nucleon Dirac spinor, $P=(p_i+p_f)/2$ is the average nucleon momentum, $\Delta=p_f-p_i$ is the momentum transfer with invariant $t=\Delta^2$, $M_N$ is the nucleon mass, $\gamma^\mu$ and $\sigma^{\mu\nu}=i[\gamma^\mu,\gamma^\nu]/2$ are Dirac matrices, $g^{\mu\nu}$ is the Minkowski metric, and $\{\mu\nu\}$ denotes symmetrization with the trace removed. The gluon form factor $B_g$ is related to gluon angular momentum, while $\bar C_g$ describes trace and higher-order EMT contributions.

Using the prescriptions of Refs.~\cite{HallC1,HallC2,HallCGFF,CLAS}, holographic QCD and GPD-based models were used to extract the $A_g(t)$ and $C_g(t)$ mechanical form factors by fitting the $t$-differential cross sections. All assumptions described below, except those concerning final-state interactions and $m_C$, were previously applied in the free-proton extractions, and the same procedure was followed here for the bound neutron and proton to enable direct comparison with the free-proton results. The primary assumption is a dominant two-gluon exchange production mechanism, which connects the \Jp photoproduction cross section to the gluon structure of the nucleon. Since lattice QCD predicts a small $B_g(t)$~\cite{LQCD2,LQCD3}, it is neglected. The poorly constrained $\bar C_g(t)$ is also neglected, although some predictions indicate it may be sizable~\cite{Cbar1,Cbar2}. The form factors are parametrized using tripole functions:

\begin{equation}\label{CalcGFF}
    A_g(t)= \frac{A_g(0)}{\left( 1 - \frac{t}{m^2_A}\right)^3},~
    C_g(t)= \frac{C_g(0)}{\left( 1 - \frac{t}{m^2_C}\right)^3},
\end{equation}

\noindent
where $A_g(0)$ and $C_g(0)$ are the form factors at $t=0$, and $m_A$ and $m_C$ determine their $t$ dependence. For $A_g(t)$, $A_g(0)$ was constrained to the CT18 parton distribution function result of $0.414\pm0.008$~\cite{CT18}. For $C_g(t)$, $m_C$ was constrained to the value extracted from free-proton measurements~\cite{CLAS}, following an analysis that demonstrated its invariance as a function of $E_{\gamma}$. Rather than fixing these parameters to their central values, they were included as Gaussian constraints, allowing variation within their uncertainties from CT18 and the free-proton data. final-state interaction contributions were neglected as stated previously.

\begin{figure}[hbtp]
    \centering   
    \includegraphics[width=\linewidth]{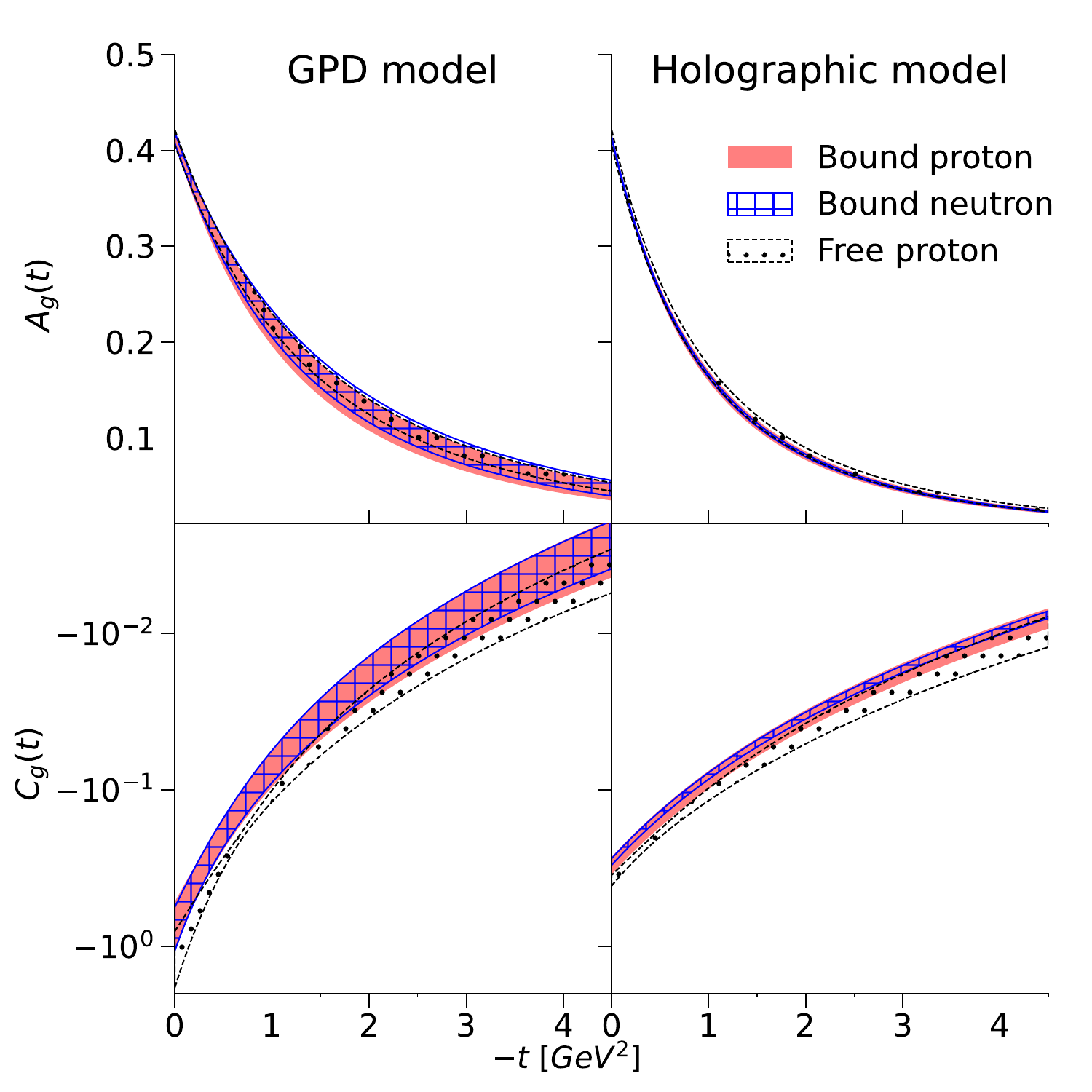}
    \caption[The $A_g(t)$ and $C_g(t)$ gravitational form factors in the GPD~\cite{GPD2} and holographic QCD~\cite{Holo4}-based models.]{The $A_g(t)$ and $C_g(t)$ gravitational form factors obtained by fitting the GPD~\cite{GPD2} (left) and holographic QCD~\cite{Holo4}-based (right) models to the free proton (dashed black), bound proton (red shaded), and bound neutron (hashed blue) data. The free proton curves are obtained from the parameters reported in Ref.~\cite{CLAS}.}
    \label{art:fig_GFFs}
\end{figure}

The holographic QCD~\cite{Holo4} and GPD~\cite{GPD2}-based models were fitted to the bound neutron and proton differential cross sections in the same way as the free-proton data in Ref.~\cite{CLAS}. The extracted $A_g(t)$ and $C_g(t)$ form factors are shown in Fig.~\ref{art:fig_GFFs}. The fitted parameters and correlations are given in Table~\ref{itab_GFFparams}. The mass and scalar radii extracted from these form factors are shown in Table~\ref{art:tab_mrGFF}. These radii characterize the spatial distribution of gluonic energy and the distribution of the scalar energy density associated with the trace of the EMT, respectively. The neutron and proton cross sections were then combined to determine the bound nucleon form factors and radii.

\begin{table}[htbp]
\centering
\begin{ruledtabular}
\begin{tabular}{clccc}
Radius & Target & {VMD} & {GPD} & {Holo.} \\\hline
 & Bound Proton  & 0.48$\pm$0.08 & $0.83\pm0.11$ & $0.67\pm0.01$ \\
Mass & Bound Neutron & 0.48$\pm$0.07 & $0.83\pm0.10$ & $0.66\pm0.01$ \\
Radius & Bound Nucleon & 0.48$\pm$0.05 & $0.81\pm0.08$ & $0.67\pm0.01$ \\
~[fm] & Free Proton & 0.52$\pm$0.02 & $1.00\pm0.17$ & $0.70\pm0.02$ \\
\hline
& Bound Proton & N/A & $1.31\pm0.20$ & $0.92\pm0.04$ \\
Scalar & Bound Neutron & N/A & $1.31\pm0.18$ & $0.90\pm0.01$ \\
Radius & Bound Nucleon & N/A & $1.28\pm0.15$ & $0.91\pm0.02$ \\
~[fm] & Free Proton & N/A & $1.64\pm0.31$ & $0.99\pm0.03$ \\
\end{tabular}
\end{ruledtabular}
\caption{Mass and scalar radii of bound and free nucleons~\cite{CLAS} extracted from fits of the VMD~\cite{MassRad}, GPD~\cite{GPD2}, and holographic QCD~\cite{Holo4}-based models to the differential cross section as a function of $t$. The free proton radii are obtained from Ref.~\cite{CLAS}.}
\label{art:tab_mrGFF}
\end{table}

Overall, the differential cross sections presented in this Letter point towards similar gluonic properties for the bound neutron and proton. Conclusions taken from this work should be strongly tempered by the assumptions detailed above and, in particular, the assumption of a dominant two-gluon exchange production mechanism that is essential in relating the \Jp photoproduction cross section to the gluon structure of the nucleon. Applying Gaussian constraints to the fit parameters also reduces the uncertainty on the extracted mechanical form factors. Within these assumptions, the $A_g(t)$ and $C_g(t)$ form factors for the bound neutron and proton are in good agreement, as are the mass and scalar radii, as shown in Fig.~\ref{art:fig_GFFs} and Table~\ref{art:tab_mrGFF}. The absolute value of the $C_g(t)$ form factor appears to be smaller for the bound neutron and proton than for the free proton, which is reflected by the mass and scalar radii of the bound nucleons that are found to be smaller than that of the free proton. Reference~\cite{SRCCT} explained a potential increase in the sub-threshold \Jp cross section as compared to the mean field expectation either with an increase in the overall magnitude of the cross section or a modification of the $t$-dependence of the cross section. The latter hypothesis is in line with the measurements presented in this Letter that imply a reduction in the effective gluon radius of bound nucleons.

\section{Conclusions}
\label{sec:Conclusions}

In this Letter, data collected with the CLAS12 experiment at JLab have been analyzed to provide the first measurement of near-threshold \Jp photoproduction on the bound neutron. The total and $t$-differential cross sections were also obtained for the bound proton. Agreement between the bound neutron and proton cross sections indicates that any isospin-breaking effects are smaller than the experimental uncertainties. The comparison of the \Jp cross sections on the neutron and proton provides further constraints on the open-charm contributions to the \Jp cross section. Holographic QCD and GPD-based models were used to estimate the $A_g(t)$ and $C_g(t)$ mechanical form factors of the bound neutron and bound proton. These are in good agreement, as are the mass and scalar radii obtained from the form factors. Finally, assuming a dominant two-gluon exchange mechanism and within the model assumptions underlying the extraction, these data suggest smaller mass and scalar radii for the bound nucleon than for the free nucleon, consistent with previous measurements of sub-threshold \Jp production on the bound proton and suggestive of medium modifications to the nucleon’s gluon structure.

The data analyzed in this work were taken with roughly 40\% of the approved run time, with the remaining data to be taken in coming years. Improvements to the CLAS12 post-processing framework will also enable taking the remaining data at increased luminosities, resulting in an estimated factor of two decrease in the statistical uncertainties reported here. The \Jp decay channel to a di-muon pair will also be investigated. Additional data has been taken on a range of heavier nuclear targets: carbon, aluminum, copper, tin, and lead. These data are currently being exploited towards studying potential deformations of the gluon structure of the bound nucleon due to the nuclear environment. Together with the present measurement, these data will provide benchmarks for the \Jp program at the Electron-Ion Collider, where precise nuclear measurements will further probe gluonic structure and its modification in nuclei~\cite{EIC}.

\section*{Acknowledgments}
The authors acknowledge the outstanding efforts of the JLab staff that made this experiment possible. 

This work was supported in part by the U.S. Department of Energy, the National Science Foundation (NSF), and the Italian Istituto Nazionale di Fisica Nucleare (INFN), the French Centre National de la Recherche Scientifique (CNRS), the French Commissariat à l’Energie Atomique (CEA), the UK Science and Technology Facilities Council (STFC), the National Research Foundation of Korea (NRF), the Helmholtz-Forschungsakademie
Hessen für FAIR (HFHF), the Chilean Agencia Nacional de Investigacion y Desarollo (ANID), the Scottish Universities Physics Alliance (SUPA), the Skobeltsyn Nuclear Physics Institute and Physics Department at the Lomonosov Moscow State University. This work was
also supported by the Marie Sklodowska-Curie Grant Agreement No.~101003460. This material is based upon work supported by the U.S. Department of Energy, Office of Science, Office of Nuclear Physics under Contract No.~89243126CSC000213.


\bibliographystyle{apsrev4-1} 
\bibliography{Article_bibliography}

\appendix{}

\newpage
\section{End Matter}

\subsection{Numerical Results}
\label{Res}

Table~\ref{tab_toteep} contains the \Jp total cross section as a function of $E_{\gamma}$ produced on the bound proton.
\begin{table}[h!]
    \centering
    \begin{ruledtabular}
    \begin{tabular}{cc}
    \textbf{$\langle E_{\gamma} \rangle$} {[GeV]}& \textbf{$\sigma$} {[nb]}\\
    \hline
    \noalign{\vskip 2pt}
    8.75$^{+0.15}$$_{-0.24}$ & 0.11 $\pm$ 0.03 $\pm$ 0.03 \\
    9.16$^{+0.09}$$_{-0.11}$ & 0.23 $\pm$ 0.08 $\pm$ 0.05 \\
    9.40$^{+0.07}$$_{-0.07}$ & 0.52 $\pm$ 0.11 $\pm$ 0.11 \\
    9.70$^{+0.14}$$_{-0.13}$ & 0.66 $\pm$ 0.15 $\pm$ 0.15 \\
    10.03$^{+0.11}$$_{-0.09}$ & 0.86 $\pm$ 0.17 $\pm$ 0.17 \\
    10.32$^{+0.12}$$_{-0.08}$ & 0.92 $\pm$ 0.25 $\pm$ 0.18 \\
    \end{tabular}
    \end{ruledtabular}
    \caption[The \Jp total cross section as a function of $E_{\gamma}$ produced on the bound proton.]{The \Jp total cross section as a function of $E_{\gamma}$ produced on the bound proton. The cross section is quoted with statistical and systematic uncertainties.}
    \label{tab_toteep}
\end{table}

Table~\ref{tab_toteen} contains the \Jp total cross section as a function of $E_{\gamma}$ produced on the bound neutron.
\begin{table}[h!]
    \centering
    \begin{ruledtabular}
    \begin{tabular}{cc}
    \textbf{$\langle E_{\gamma} \rangle$} {[GeV]}& \textbf{$\sigma$} {[nb]}\\
    \hline
    \noalign{\vskip 2pt}
    8.51$^{+0.16}$$_{-0.21}$ & 0.13 $\pm$ 0.03 $\pm$ 0.03 \\
    8.85$^{+0.07}$$_{-0.06}$ & 0.36 $\pm$ 0.15 $\pm$ 0.07 \\
    9.08$^{+0.08}$$_{-0.09}$ & 0.41 $\pm$ 0.13 $\pm$ 0.08 \\
    9.41$^{+0.13}$$_{-0.14}$ & 0.62 $\pm$ 0.15 $\pm$ 0.12 \\
    9.92$^{+0.28}$$_{-0.20}$ & 0.84 $\pm$ 0.19 $\pm$ 0.16 \\
    \end{tabular}
    \end{ruledtabular}
    \caption[The \Jp total cross section as a function of $E_{\gamma}$ produced on the bound neutron.]{The \Jp total cross section as a function of $E_{\gamma}$ produced on the bound neutron. The cross section is quoted with statistical and systematic uncertainties.}
    \label{tab_toteen}
\end{table}

Table~\ref{tab_difeep} contains the \Jp $t$-differential cross section produced on the the bound proton.
\begin{table}[h!]
    \centering
    \begin{ruledtabular}
    \begin{tabular}{ccc}
    \textbf{$\langle -t \rangle$} {[GeV$^2$]}& \textbf{$\langle E_{\gamma} \rangle$} {[GeV]}& \textbf{$d\sigma/dt$}  {[nb/GeV$^2$]} \\
    \hline
    \noalign{\vskip 2pt}
    0.50$^{+0.13}$$_{-0.13}$ & 9.14$^{+0.64}$$_{-0.57}$ & 0.70 $\pm$ 0.30 $\pm$ 0.16 \\
    0.84$^{+0.11}$$_{-0.10}$ & 9.19$^{+0.54}$$_{-0.58}$ & 0.39 $\pm$ 0.06 $\pm$ 0.09 \\
    1.23$^{+0.18}$$_{-0.16}$ & 9.22$^{+0.50}$$_{-0.58}$ & 0.24 $\pm$ 0.04 $\pm$ 0.05 \\
    1.77$^{+0.21}$$_{-0.19}$ & 9.24$^{+0.51}$$_{-0.60}$ & 0.13 $\pm$ 0.02 $\pm$ 0.04 \\
    2.33$^{+0.19}$$_{-0.15}$ & 9.25$^{+0.48}$$_{-0.64}$ & 0.09 $\pm$ 0.02 $\pm$ 0.02 \\
    3.15$^{+0.55}$$_{-0.36}$ & 9.25$^{+0.49}$$_{-0.61}$ & 0.03 $\pm$ 0.02 $\pm$ 0.01 \\
    \end{tabular}
    \end{ruledtabular}
    \caption[The \Jp $t$-differential cross section produced on the bound proton.]{The \Jp $t$-differential cross section produced on the bound proton. The cross section is quoted with statistical and systematic uncertainties.}
    \label{tab_difeep}
\end{table}

Table~\ref{tab_difeen} contains the \Jp $t$-differential cross section produced on the the bound neutron.
\begin{table}[h!]
    \centering
    \begin{ruledtabular}
    \begin{tabular}{ccc}
    \textbf{$\langle -t \rangle$}  {[GeV$^2$]}& \textbf{$\langle E_{\gamma} \rangle$} {[GeV]}& \textbf{$d\sigma/dt$} {[nb/GeV$^2$]} \\
    \hline
    \noalign{\vskip 2pt}
    0.56$^{+0.09}$$_{-0.09}$ & 9.06$^{+0.67}$$_{-0.55}$ & 0.59 $\pm$ 0.26 $\pm$ 0.15 \\
    0.85$^{+0.10}$$_{-0.11}$ & 9.17$^{+0.60}$$_{-0.59}$ & 0.42 $\pm$ 0.15 $\pm$ 0.09 \\
    1.23$^{+0.18}$$_{-0.15}$ & 9.18$^{+0.56}$$_{-0.59}$ & 0.25 $\pm$ 0.10 $\pm$ 0.06 \\
    1.72$^{+0.19}$$_{-0.14}$ & 9.21$^{+0.51}$$_{-0.61}$ & 0.16 $\pm$ 0.05 $\pm$ 0.04 \\
    2.74$^{+0.78}$$_{-0.45}$ & 9.20$^{+0.56}$$_{-0.63}$ & 0.05 $\pm$ 0.01 $\pm$ 0.01 \\
    \end{tabular}
    \end{ruledtabular}
    \caption[The \Jp $t$-differential cross section produced on the bound neutron.]{The \Jp $t$-differential cross section produced on the bound neutron. The cross section is quoted with statistical and systematic uncertainties.}
    \label{tab_difeen}
\end{table}

Table~\ref{itab_GFFparams} summarizes the complete set of fit results for the bound nucleon mechanical form factors in the GPD and holographic QCD parameterizations. In addition to the best-fit parameter values and fit quality, the table lists the parameter correlation coefficients required to reconstruct the full covariance matrix. Together, these quantities provide all information necessary to reproduce the form factor parameterizations, propagate uncertainties to derived observables, and calculate the corresponding mass and scalar radii.

\begin{table}[h!]
\centering
\begin{ruledtabular}
\begin{tabular}{clcc}
& Target & {GPD} & {Holo.} \\
\hline
& Proton  & 0.85 & 0.11 \\
\textbf{$\chi^2/\mathrm{ndf}$} & Neutron & 0.53 & 0.01 \\
& Nucleon & 0.59 & 0.06 \\
\hline
& Proton  & $0.41\pm0.01$ & $0.41\pm0.01$ \\
\textbf{$A_0^\ast$} & Neutron & $0.41\pm0.01$ & $0.41\pm0.01$ \\
& Nucleon & $0.41\pm0.01$ & $0.41\pm0.01$ \\
\hline
& Proton  & $-0.80\pm0.28$ & $-0.31\pm0.04$ \\
\textbf{$C(0)$} & Neutron & $-0.82\pm0.25$ & $-0.29\pm0.01$ \\
& Nucleon & $-0.76\pm0.20$ & $-0.30\pm0.02$ \\
\hline
& Proton  & $2.01\pm0.13$ & $1.67\pm0.04$ \\
\textbf{$m_A$} & Neutron & $2.06\pm0.11$ & $1.67\pm0.01$ \\
& Nucleon & $2.05\pm0.09$ & $1.67\pm0.02$ \\
\hline
& Proton  & $0.92\pm0.11$ & $1.38\pm0.03$ \\
\textbf{$m_C^\ast$} & Neutron & $0.90\pm0.09$ & $1.38\pm0.01$ \\
& Nucleon & $0.94\pm0.09$ & $1.38\pm0.02$ \\
\hline
& Proton  & $-0.10$ & $0.00$ \\
\textbf{$\rho(m_A,A_0)$} & Neutron & $-0.09$ & $0.00$ \\
& Nucleon & $-0.12$ & $0.00$ \\
\hline
& Proton  & $-0.07$ & $-0.05$ \\
\textbf{$\rho(A_0,C(0))$} & Neutron & $-0.07$ & $-0.04$ \\
& Nucleon & $-0.08$ & $-0.06$ \\
\hline
& Proton  & $-0.13$ & $-0.85$ \\
\textbf{$\rho(m_A,C(0))$} & Neutron & $-0.21$ & $-0.86$ \\
& Nucleon & $+0.08$ & $-0.79$ \\
\hline
& Proton  & $+0.44$ & $+0.29$ \\
\textbf{$\rho(m_A,m_C)$} & Neutron & $+0.37$ & $+0.24$ \\
& Nucleon & $+0.54$ & $+0.37$ \\
\hline
& Proton  & $+0.76$ & $+0.11$ \\
\textbf{$\rho(m_C,C(0))$} & Neutron & $+0.73$ & $+0.08$ \\
& Nucleon & $+0.83$ & $+0.14$ \\
\end{tabular}
\end{ruledtabular}
\caption[Fitted parameters for bound nucleon mechanical form factors in the GPD and holographic QCD-based models.]{Complete fit results for the bound nucleon mechanical form factors in the GPD and holographic QCD parameterizations. The table lists the goodness of fit ($\chi^2/\mathrm{ndf}$), best-fit parameter values with statistical uncertainties, and the parameter correlation coefficients defining the covariance matrix used for uncertainty propagation. The Gaussian-constrained parameters are denoted by $^\ast$.}
\label{itab_GFFparams}
\end{table}
\end{document}